\documentclass[aps,prb,twocolumn,superscriptaddress,floatfix,showpacs]{revtex4}
\usepackage{graphicx}
\usepackage{amsmath}
\usepackage{graphicx}

\begin{document}

\title{Renormalization effects in interacting quantum dots coupled to superconducting leads}

\author{David Futterer}
\affiliation{Theoretische Physik, Universit\"at Duisburg-Essen and CENIDE, 47048 Duisburg, Germany}
 
 \author{Jacek Swiebodzinski}
\affiliation{Theoretische Physik, Universit\"at Duisburg-Essen and CENIDE, 47048 Duisburg, Germany}
 
\author{Michele Governale}
\affiliation{School of Chemical and Physical Sciences and MacDiarmid Institute for Advanced Materials
and Nanotechnology, Victoria University of Wellington, PO Box 600, Wellington 6140, New Zealand}

\author{J\"urgen K\"onig}
\affiliation{Theoretische Physik, Universit\"at Duisburg-Essen and CENIDE, 47048 Duisburg, Germany}

\date{\today}

\begin{abstract}
We study subgap transport through an interacting quantum dot tunnel coupled to one normal and two superconducting leads.
To check the reliability of an approximation of an infinitely-large gap $\Delta$ in the superconducting leads and weak tunnel coupling to the normal lead, we perform a $1/\Delta$ expansion, and we analyze next-to-leading order corrections in the tunnel coupling to the normal lead.
Furthermore, we propose a resummation approach to calculate the Andreev bound states for finite $\Delta$.
The results are substantially more accurate than those obtained by mean-field treatments and favorably compare with numerically exact results.
\end{abstract}

%74.45.+c       Proximity effects; Andreev effect; SN and SNS junctions
%73.63.Kv      Quantum dots

\maketitle

\section{Introduction}
The tunnel coupling to superconducting reservoirs can induce superconducting correlations in quantum dots.\cite{franceschi10,martin-rodero11}
Interesting linear- and nonlinear transport effects arising from Andreev reflection\cite{degennes63,andreev64} can be studied in hybrid normal-superconducting systems comprising quantum dots  in the presence of strong Coulomb interaction.
During the past years, there has been tremendous experimental progress in coupling quantum dots to superconductors.\cite{buitelaar02,buitelaar03,cleuziou06,jarillo-herrero06,ingerslev06,grove-rasmussen07,ingerslev07,eichler07,sand-jespersen07,eichler09,deacon10,deacon10prb,vandam06,hofstetter09,herrmann10,das12,pillet10,dirks11}
Prominent examples include the observation of $0-\pi$~transitions in a Josephson junction\cite{vandam06} and splitting of Cooper pairs.\cite{hofstetter09,herrmann10,das12}

An exact theoretical treatment of the transport properties of such devices requires to account for Coulomb interaction, superconducting correlations, and non-equilibrium at the same time.
To circumvent this challenge, various approximation schemes have been proposed.
In the case of vanishing Coulomb interaction, an exact solution can be obtained within a scattering approach.\cite{beenakker95}
As a first step beyond, Coulomb interaction has been treated perturbatively,\cite{fazio98,schwab99,clerk00,cuevas01,avishai03,dell-anna08,koerting10} which, however, is not justified for the large charging energies that are typical for small quantum dots.
An alternative possibility is to allow for strong Coulomb interaction but perform a perturbation expansion in the strength of the tunnel coupling between dot and leads.\cite{pala07,governale08}
However,  in the limit of an infinitely-large superconducting gap, an exact treatment of both  the Coulomb interaction and the tunnel coupling between quantum dot and superconducting leads is possible.
\cite{governale08,rozhkov00,tanaka07,karrasch08}
In this case, quasiparticle tunneling is completely suppressed and transport from and to the superconducting leads is fully sustained by Cooper pairs. 
The formation of Andreev-bound states (ABSs) in the quantum dot indicates that superconducting correlations are induced via the proximity to superconducting leads.
The dependence of the ABS energies on the quantum dot's level position is directly reflected in the transport spectrum of the system.

A  diagrammatic real-time approach to nonequilibrium transport through quantum dots coupled both to normal and superconducting leads has been introduced in Ref.~\onlinecite{governale08}. 
This method has been applied in the limit of an infinitely-large superconducting gap, $\Delta \rightarrow \infty$, to study a variety of transport phenomena in different quantum-dot setups, \cite{futterer09,sothmann10,futterer10,braggio11,eldridge10,hiltscher11,moghaddam12,hiltscher12} 
including, for example,
pure spin-current generation,\cite{futterer10} shot-noise suppression,\cite{braggio11} Cooper-pair splitting,\cite{eldridge10,hiltscher11} and time-dependent driving.\cite{moghaddam12}
In this limit, both the charging energy and the coupling to the superconducting leads can be treated nonperturbatively by the diagrammatic technique and a perturbation expansion is only performed in the coupling to the normal-conducting reservoirs.

The calculations in the infinite-gap limit neglect quasiparticle contributions to transport.
They, furthermore, approximate the ABS energies since the latter are a function of the superconducting gap.
In the experiments, the superconducting gap $\Delta$ is often of the same order of magnitude as other energy scales such as charging energy or tunnel-coupling strength.
For this reason, it is quite natural to investigate the quality of the approximation introduced by considering the infinite-gap limit.
This is the main goal of the present paper.
Furthermore, we analyze the effect of next-to-lowest order corrections in the tunnel coupling to the normal leads.

A finite superconducting gap in the leads affects the Andreev and Josephson transport through quantum dots in two ways. 
First, it modifies the ABSs. 
This affects the position of bias or gate voltages at which a new transport channel is opened.
Second, the value of the current away from these threshold voltages is changed.
To address the effect of a finite $\Delta$, we start by performing a systematic expansion in $1/\Delta$ around $\Delta\rightarrow\infty$.
Exemplarily, we will focus on a quantum dot tunnel coupled to one normal and two superconducting reservoirs and compare the currents obtained by the $1/\Delta$ expansion with the $\Delta\rightarrow\infty$ limit.\cite{governale08}
Since we only focus on the DC Josephson current we assume the two superconductors to be on the same chemical potential and, thus, the subgap structure expected from multiple Andreev reflection processes\cite{yeyati97,johansson99,sun02,yeyati03,dell-anna08,jonckheere09} does not occur. 
We find that away from the threshold voltages, both the Andreev and the Josephson currents are hardly modified, i.e., the $\Delta\rightarrow\infty$ limit provides quite an accurate description of subgap transport.
The $1/\Delta$ expansion, furthermore, indicates a renormalization of the ABS energies.
It can, however, only predict the direction towards which the energies are shifted.
To determine the position of the ABS for finite values of $\Delta$ we perform, as an alternative approximation scheme, a partial resummation of diagrams.
The results within this approach are considerably more accurate than those obtained by
a simple Hartree-Fock treatment and  compare favorably with the numerical renormalization group (NRG) data of Mart\'in-Rodero and Levy Yeyati.\cite{martin-rodero12}

Finally, we present (for the $\Delta\rightarrow\infty$ limit) results from a systematic perturbation expansion to next-to-lowest order in the tunnel coupling to the normal leads.
In this way, we go beyond the regime of weak tunnel coupling but do not cover Kondo correlations.\cite{deacon10,deacon10prb,graeber04,hofstetter10,fazio98,schwab99,cho99,clerk00,cuevas01,sun01,aono03,siano04,krawiec04,splettstoesser07,tanaka07-2,domanski08,domanski08-2,yamada10,yamada11}
We find that the next-to-leading-order correction leads to a renormalization of the gate-voltage position at which a $0-\pi$~transition occurs in the Josephson current.
This renormalization also affects the Andreev current and the average quantum-dot charge.
 
\section{Model and Method}
\label{sec:model}

\subsection{Hamiltonian}

\begin{figure}
\includegraphics[width=0.9\columnwidth]{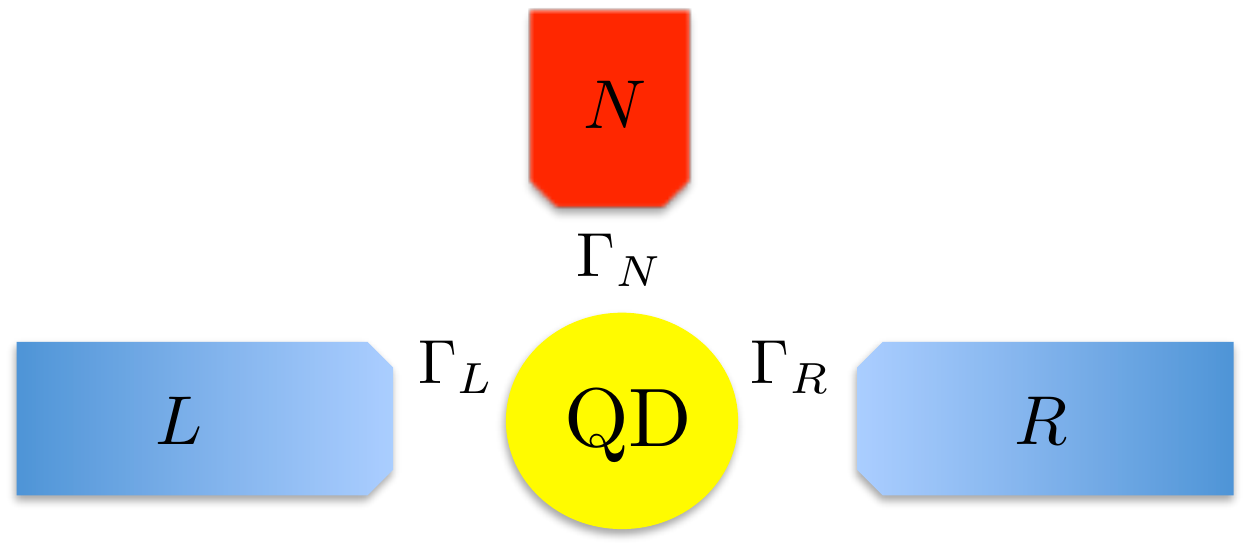}
\caption{(Color online) Setup: a quantum dot is tunnel coupled
to one normal and two superconducting leads.
\label{setup}}
\end{figure}
We focus on a three-terminal setup composed of a normal conductor and two superconductors
tunnel coupled to an interacting single-level quantum dot, see Fig.~\ref{setup}. The system's Hamiltonian is
given by $H=H_{\rm{dot}}+\sum_{\eta}H_{\eta}+H_{\rm{tunn,\eta}}$, with $\eta \in \{ L,R,N \}$.
We assume the normal lead to be a reservoir of noninteracting electrons and model the superconductors
with the mean-field BCS Hamiltonian
\begin{equation}
H_{\eta}=\sum_{k \sigma} \epsilon_{\eta k}
c_{\eta k \sigma}^\dagger c_{\eta k \sigma}- \delta_{\eta, S} \sum_{k}\left(\Delta e^{i \Phi_{\eta}}  c_{\eta -k \downarrow} 
c_{\eta k \uparrow}+\text{H.c.}\right), 
\end{equation}
with $\epsilon_{\eta k}$ being the single-particle energies and  $c_{\eta k \sigma}^{(\dagger)}$ are the
annihilation (creation) operators for electrons in lead $\eta$ with momentum $k$ and spin $\sigma$. Here, $\Delta$ is the modulus of the superconducting pair potential, which we assume to be the same in both superconductors, and $\Phi_{\eta}$ is the corresponding phase in lead $\eta=S$ with $S \in\{ L,R\}$.
We measure all occurring energies with respect to the Fermi level of the superconductors, i.e. $\mu_{L,R}=0$.
Furthermore, we assume all occurring excitation energies of the dot to lie inside the superconducting gap of the leads so that only subgap transport takes place.
  
The Anderson Hamiltonian describing the single-level quantum dot reads
\begin{equation} 
\label{AndersonHamiltonian}
H_{\rm{dot}}=\sum_{\sigma} \epsilon d_{\sigma}^{\dagger} d_{\sigma} + U n_{\uparrow} 
n_{\downarrow},
\end{equation}
where $\epsilon$ is  the energy of the spin-degenerate single-particle level, and $U$ the on-site Coulomb repulsion. 
The dot's annihilation (creation) operators of spin $\sigma$ are given by $d_{\sigma}^{(\dagger)}$, and
$n_{\sigma}=d_{\sigma}^{\dagger} d_{\sigma} $ is the corresponding number operator. Tunneling between
dot and leads is described by the tunneling Hamiltonian
\begin{equation}
  H_{\rm{tunn},\eta}= 
  V_{\eta} \sum_{k \sigma} \left( c_{\eta k \sigma}^\dagger d_\sigma +
  {\rm H.c.} \right), 
\label{htun}
\end{equation}
with the spin and momentum independent tunnel matrix elements $V_{\eta}$. We define the tunnel-coupling strengths
as $\Gamma_{\eta}=2 \pi |V_{\eta}|^2 \rho_{\eta}$, where $\rho_{\eta}$ is the density of states in lead $\eta$
that we assume to be independent of spin and energy.

\subsection{Diagrammatic Technique}

\subsubsection{Effective Dot Hamiltonian}
The main idea of the diagrammatic real-time technique is to integrate out the leads' degrees of freedom to arrive at a reduced density matrix for the quantum dot.
As a basis for the reduced density matrix one may use the eigenstates of the decoupled quantum dot, $H_{\rm{dot}}$.
This is a convenient choice for weak tunnel couplings to the superconductors.\cite{pala07}
In the opposite limit of strong  coupling to the superconductor, a resummation of an infinite number of diagrams is required. This resummation can be carried out  exactly in the $\Delta\rightarrow\infty$ limit. \cite{governale08}
For a systematic $1/\Delta$-expansion, however, it is more convenient to work in the eigenbasis of $H_{\rm{eff}} = H_{\rm{dot}}-H_{\rm{p}}$, with the pairing Hamiltonian $H_{\rm{p}}=\chi^{*} d_{\downarrow} d_{\uparrow} + \chi d_{\uparrow}^{\dagger} d_{\downarrow}^{\dagger}$ and $\chi=\frac{1}{2}\sum_{\eta=L,R} \Gamma_{\eta} e^{i \Phi_{\eta}}$.
The remaining part of the Hamiltonian
\begin{equation}
H - H_{\rm{eff}} = H_{\rm{p}}+\sum_{\eta}H_{\eta}+H_{\rm{tunn,\eta}},
\end{equation}
is treated diagrammatically as a perturbation.
Thereby, $H_{\rm{p}}$ has been chosen such that in the $\Delta\rightarrow\infty$ limit the diagrammatic contributions stemming from $H_{\rm{p}}$ exactly cancel those from the tunneling to the superconducting leads, i.e., $H_{\rm{eff}}$
describes the hybrid system of quantum dot and superconducting leads in the limit of an infinite superconducting gap.

The eigenstates of $H_{\rm{eff}}$ are $\left|\uparrow\right\rangle$, $\left|\downarrow\right\rangle$, and
$\vert \pm \rangle = \frac{1}{\sqrt{2}} \left[ \mp e^{-i \Phi_{\chi}/2} \sqrt{1\mp\frac{\delta}{2 \epsilon_{A}}} \; \vert 0 \rangle + e^{i\Phi_{\chi}/2} \sqrt{1\pm\frac{\delta}{2\epsilon_{A}}} \; \vert d \rangle \right] $,
with eigenenergies $\epsilon$, $\epsilon$, and $E_\pm=\frac{\delta}{2} \pm \epsilon_{A}$, respectively.
Here, $\Phi_{\chi} = \mathrm{arg}(\chi)$, $\epsilon_{A} = \sqrt{(\frac{\delta}{2})^{2} + \vert \chi \vert^{2}}$, and $\delta=2 \epsilon + U$ is the detuning between the energies for empty and doubly-occupied quantum dot.
The superconducting proximity effect is indicated by the fact that $| 0 \rangle$ and $|d\rangle$ are no longer eigenstates  but appear as linear combinations in $| \pm \rangle$.
The mixing between $| 0 \rangle$ and $|d\rangle$ becomes largest around zero detuning, $\delta\sim 0$. 
 
The excitation energies associated with $H_{\rm{eff}}$ are given by the differences of the eigenenergies of states with even and odd dot occupation numbers, 
\begin{equation}
\label{ABS}
E_{A, \gamma', \gamma}^{\Delta\rightarrow\infty}= \gamma' \frac{U}{2} + \gamma \epsilon_{A},
\end{equation}
with $\gamma', \gamma = \pm$.
They are nothing but the Andreev bound state (ABS) energies\cite{footnote-ABS} in the $\Delta\rightarrow\infty$ limit.

\subsubsection{Generalized Master Equation}
The system's dynamics is determined by a generalized master equation for the reduced density matrix $\rho_{\rm{red}}$, which is obtained by integrating out the lead's degrees of freedom.
Its matrix elements $P_{\chi_2}^{\chi_1}=\langle \chi_1|\rho_{\rm{red}}|\chi_2\rangle$ obey, in the stationary limit, the generalized master equation 
\begin{equation}
\label{master-equation}
i\left( E_{\chi_1}-E_{\chi_2} \right) P_{\chi_2}^{\chi_1}=\sum_{\chi_1' \chi_2'} W_{\chi_2 \chi_2'}^{\chi_1 \chi_1'} P_{\chi_2'}^{\chi_1'},
\end{equation}
where $W_{\chi_2,\chi_2'}^{\chi_1,\chi_1'}$ are generalized rates that can be computed in a diagrammatic way, see Sec.~\ref{diagrammatic-rules}.
As explained in the previous section, we work in the basis $\{ \uparrow, \downarrow, +, - \}$.
This has the consequence that the off-diagonal matrix elements on the right hand side of Eq.~(\ref{master-equation}) vanish to lowest order in $\Gamma_N$ and $1/\Delta$ as long as $\epsilon_A \gtrsim \Gamma_N$.
Only their next-order correction remains finite.
 
Superconducting correlations in the quantum dot are associated with coherent superpositions of an empty and a doubly-occupied dot.
This motivates the following definition of an isospin,\cite{pala07,governale08}
\begin{equation}
	I_x=\frac{P_0^d+P_d^0}{2}, \quad I_y=i\frac{P_0^d-P_d^0}{2}, \quad I_z=\frac{P_d-P_0}{2} \, ,
\label{isospin_definition}
\end{equation}
formulated in the basis $\{ \uparrow, \downarrow, 0, d \}$ (the transformation to the basis $\{ \uparrow, \downarrow, +, - \}$ is straightforward).
A finite value of the isospin components $I_x$ and/or $I_y$ indicates the presence of superconducting proximity effect.

It turns out that the generalized master equations for the isospin components do not couple to the probabilities of the quantum dot to be singly occupied with either spin. 
They can be written in the form of a Bloch-like equation
\begin{equation}
	0=\frac{d\mathbf{I}}{dt}=\mathbf{A}-\mathbf{R} \cdot \mathbf{I}+\mathbf{I} \times \mathbf{B},
\end{equation}
where the terms with $\mathbf{A}$, $\mathbf{R}$, and $\mathbf{B}$ describe the generation, relaxation, and coherent rotation of the isospin.

\subsubsection{Current formulae}

The current in lead $\eta$ is given by
\begin{equation}
\label{current-formula}
	J_{\eta} =-\frac{e}{\hbar}  \sum_{\chi \chi_1' \chi_2'} W_{\chi \chi_2'}^{\chi \chi_1'  \, I_{\eta}} P_{\chi_2'}^{\chi_1'},
\end{equation}
with $W_{\chi \chi_2'}^{\chi \chi_1'  \, I_{\eta}}$ being the generalized current rates.
For a systematic perturbation expansion in powers of $\Gamma_N$ and $1/\Delta$ one needs to expand $P$, $W$, and $W^{I_\eta}$ in Eqs.~(\ref{master-equation}) and (\ref{current-formula}).

We refer to the current flowing out of the normal lead as the Andreev current and to the difference of the currents between left and right superconductor as the Josephson current.
In the limit $\Delta\rightarrow\infty$, the Andreev current and the Josephson current are directly connected to the isospin via \cite{governale08}
\begin{eqnarray}
\label{josephson_iso}
	J_{\rm{jos}}^{\Delta\rightarrow\infty}&=&\frac{2e}{\hbar}\Gamma_S I_x^{\Delta\rightarrow\infty} \sin\frac{\Phi}{2},\\
	J_{\rm{and}}^{\Delta\rightarrow\infty}&=&-\frac{4e}{\hbar}\Gamma_S I_y^{\Delta\rightarrow\infty} \cos\frac{\Phi}{2}.
\label{andreev_iso}
\end{eqnarray} 
For the $1/\Delta$ corrections to the currents, however, one needs to use the more general Eq.~(\ref{current-formula}).

\subsubsection{Diagrammatic Rules}
\label{diagrammatic-rules}
Starting from Refs.~\onlinecite{governale08} and \onlinecite{sothmann10} we derive the rules for calculating the generalized rates and the generalized current rates in the basis of the effective Hamiltonian.
The diagrammatic expansion is performed with respect to the tunneling parts of the Hamiltonian and, in addition, $H_P$.
In the following, we assume the two superconductors being at zero chemical potential, i.e. $\mu_L=\mu_R=0$ and a symmetric tunneling coupling, $\Gamma_L=\Gamma_R=\Gamma_S$.\\

(1) Draw all topologically different diagrams with fixed ordering of the vertices with respect to (real) time. The vertices may be bullets (representing tunneling) or crosses (for $H_p$). The bullets are connected in pairs by tunneling lines carrying energy $\omega_{i}$. The tunneling lines can be normal or anomalous. For each anomalous line, choose the direction (forward or backward with respect to the Keldysh contour) arbitrarily.
\\
(2) For each vertical cut between two vertices, assign a factor $1/(\Delta E+i 0^{+})$, where $\Delta E$ is the difference between the left-going and the right-going energies, including the energies of the dot states $E_{\chi}$ and the tunneling lines $\omega_{i}$.
\\
(3) For each tunneling line arising from the normal lead assign a factor $\frac{1}{2\pi} \Gamma_{N} f_{N}^{\pm}(\omega_{N})$ and for each tunneling line arising from a superconducting lead assign a factor $\frac{1}{2\pi} \Gamma_{S} \frac{\vert \omega_i \vert}{\sqrt{|\omega_i|^{2} - \Delta^{2}}} \Theta\left( \vert \omega_i \vert - \Delta \right) f^{\pm}(\omega_{i})$, where $f_{N}^{+}(\omega)=f^+(\omega-\mu_N)=f(\omega-\mu_N)=\{ 1 + \exp\left[\left( \omega -\mu_{N} \right) / k_{B}T \right]\}^{-1}$ and $f_{N}^{-}(\omega)=f^-(\omega-\mu_N)=1-f(\omega-\mu_N)$.
The upper (lower) sign applies for lines going backward (forward) with respect to the Keldysh contour. For anomalous lines, multiply an additional factor $\pm \mathrm{sign} \left(\omega_{i}  \right) \frac{\Delta}{\vert \omega_{i} \vert}$. Moreover, assign a factor $e^{i\Phi_{\chi}} \, e^{-i\Phi_{\eta}}$ for an outgoing and $e^{-i\Phi_{\chi}} \, e^{i\Phi_{\eta}}$ for an incoming anomalous line, where $\Phi_{\chi} = \mathrm{arg} \left( \chi \right)$ and $\chi = \frac{1}{2}\Gamma_S\sum_{\eta \in S}  e^{i \Phi_{\eta}}$.
\\
(4) Each bullet vertex connects a dot state $\vert \alpha \rangle$, with $\alpha\in\{ +,- \}$, to another dot state $\vert \sigma \rangle$, with $\sigma \in \{ \uparrow, \downarrow \}$, and to a tunneling line describing a tunneling electron of spin $\sigma' \in \{ \uparrow, \downarrow \}$. For each bullet vertex assign a factor $\sqrt{1-S_{\alpha} S_{\sigma} S_{\sigma'} \frac{\delta}{2 \epsilon_{A}}}/\sqrt{2}$, where $S_{\alpha=+}=S_{\sigma=\uparrow}=S_{\sigma'=\uparrow}=1$ and $S_{\alpha=-}=S_{\sigma=\downarrow}=S_{\sigma'=\downarrow}=-1$.
\\
(5) Each cross vertex connects a state $\vert \alpha \rangle$ with $\alpha \in \{ +,- \}$ to a state $\vert \alpha' \rangle$ with $\alpha' \in \{ +,- \}$. For each cross vertex with $\alpha = \alpha'$ assign a factor $(-1) S_{\alpha} \sqrt{1-\frac{\delta^{2}}{4\epsilon_{A}^{2}}} \vert \chi \vert$ and for $\alpha \neq \alpha'$ assign a factor $\frac{\delta}{2\epsilon_{A}} \vert \chi \vert$.
\\
(6) Assign an overall prefactor $-i$.\\
Furthermore, assign a factor $-1$ for each\\
(a) vertex on the lower propagator;\\
 (b) crossing of tunneling lines;\\
 (c) outgoing (incoming) bullet vertex ending (starting) in the state $\vert + \rangle$ or $\vert \uparrow \rangle$;\\
 (d) outgoing (incoming) anomalous tunneling line, in which the earlier (later) tunnel vertex with respect to the Keldysh contour is connected to a tunneling spin-up electron.
 \\
 (7) For each tunneling line, integrate over the energy $\omega_{i}$. Sum over all diagrams.\\
\\
 The generalized current rates $W_{\chi\chi'_{2}}^{\chi\chi'_{1}\, I_{\eta}}$ are evaluated in the following way:\\
 (8) Multiply each diagram that contributes to the corresponding generalized rate $W_{\chi\chi'_{2}}^{\chi\chi'_{1}}$ and where the rightmost line is associated with lead $\eta$ with a factor:\\
(a) for the rightmost line being a normal line: 1 if the line is going from the lower to the upper propagator, -1 if it is going from the upper to the lower propagator, and 0 otherwise;\\
(b) for the rightmost line being an anomalous line: 1 for incoming lines within the upper and outgoing lines within the lower propagator, $-1$ for outgoing lines within the upper and incoming lines within the lower propagator, and 0 otherwise.\\

\begin{figure}
\includegraphics[width=0.45\columnwidth]{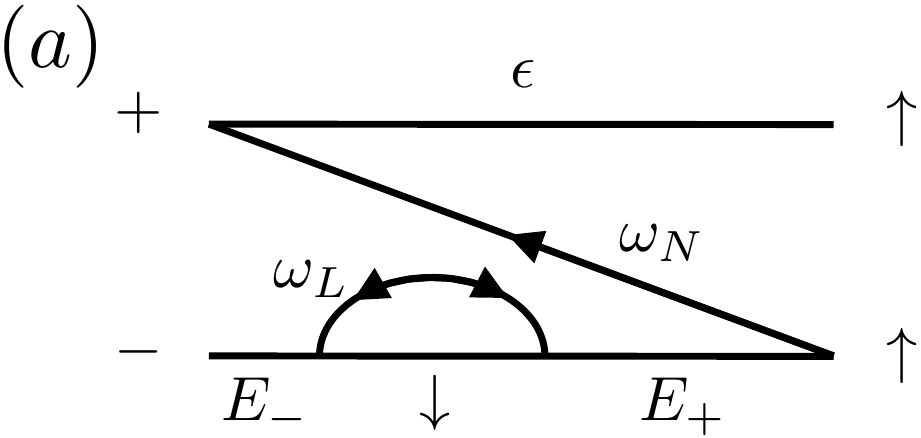}
$\quad$
\includegraphics[width=0.45\columnwidth]{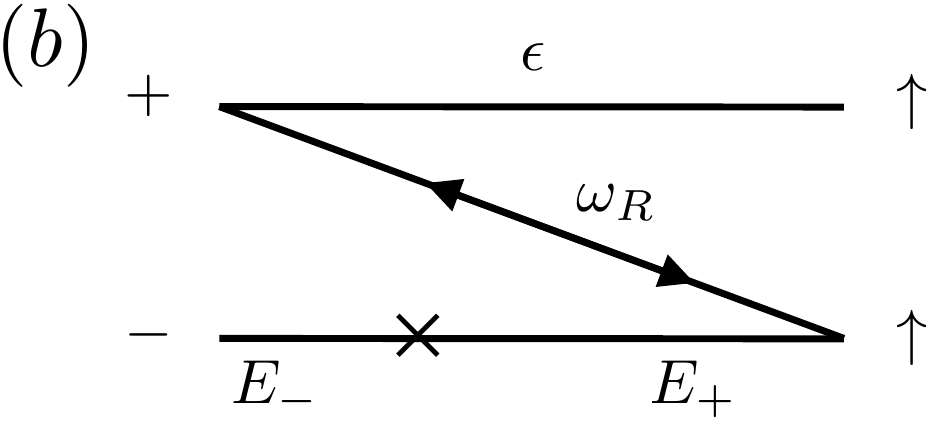}
	\caption{Example diagrams contributing to the rate $W_{\uparrow -}^{\uparrow +}$.
\label{example-diagrams}}
\end{figure}

We want to remark that, in the infinite-gap limit all terms arising from rule~(5) are exactly cancelled by the terms arising from the superconducting tunneling lines.
This reflects the fact that the effective Hamiltonian already includes the coupling to the superconductors in the infinite-gap limit. Thus, it completely describes the hybrid system of quantum dot and superconducting leads, so that only the coupling to the normal lead still needs to be accounted for perturbatively.
To illustrate how the graphical representation is translated into analytical expressions we show two exemplary diagrams here.
The first example is a diagram containing one line associated with the normal lead and one line associated with the left superconductor, see Fig.~\ref{example-diagrams}~(a), and it reads
\begin{eqnarray*}
	&&-i\int \frac{d\omega_L}{2\pi}\int \frac{d\omega_N}{2\pi} \Gamma_N \Gamma_S e^{i(\Phi_L-\Phi_{\chi})} f_N^+(\omega_N)f^-(\omega_L)\\
	&\times& \frac{1}{\omega_N+E_- -\epsilon+i0^+}  \frac{1}{\omega_N+\omega_L+i0^+}  \frac{1}{\omega_N+E_+ -\epsilon+i0^+}\\
	&\times&  \mathrm{sign} (\omega_L) \frac{\Delta}{\sqrt{\omega_L^2-\Delta^2}} \Theta(|\omega_L|-\Delta) \left(1-\frac{\delta}{2\epsilon_A}\right)^2.
\end{eqnarray*}
The second example is a diagram, see Fig.~\ref{example-diagrams}~(b), containing a line associated with the right superconductor and a cross vertex and it yields the following expression
\begin{eqnarray*}
	&&i\int\frac{d\omega_R}{2\pi} \Gamma_S e^{i(\Phi_R-\Phi_{\chi})} f^-(\omega_R) \frac{1}{E_- -\omega_R-\epsilon+i0^+}\\
	&\times& \frac{1}{E_+ - \omega_R-\epsilon+i0^+} \mathrm{sign}(\omega_R) \frac{\Delta}{\sqrt{\omega_R^2-\Delta^2}}\\
	&\times& \Theta(|\omega_R|-\Delta)  \frac{\delta |\chi|}{2\epsilon_A} \sqrt{1-\frac{\delta^2}{4\epsilon_A^2}}.
\end{eqnarray*}

\section{Finite-gap effects}
\label{results-finite-gap}

\subsection{$1/\Delta$ Expansion}
\label{finite-gap-expansion}

\begin{figure}
	\includegraphics[width=0.9\columnwidth]{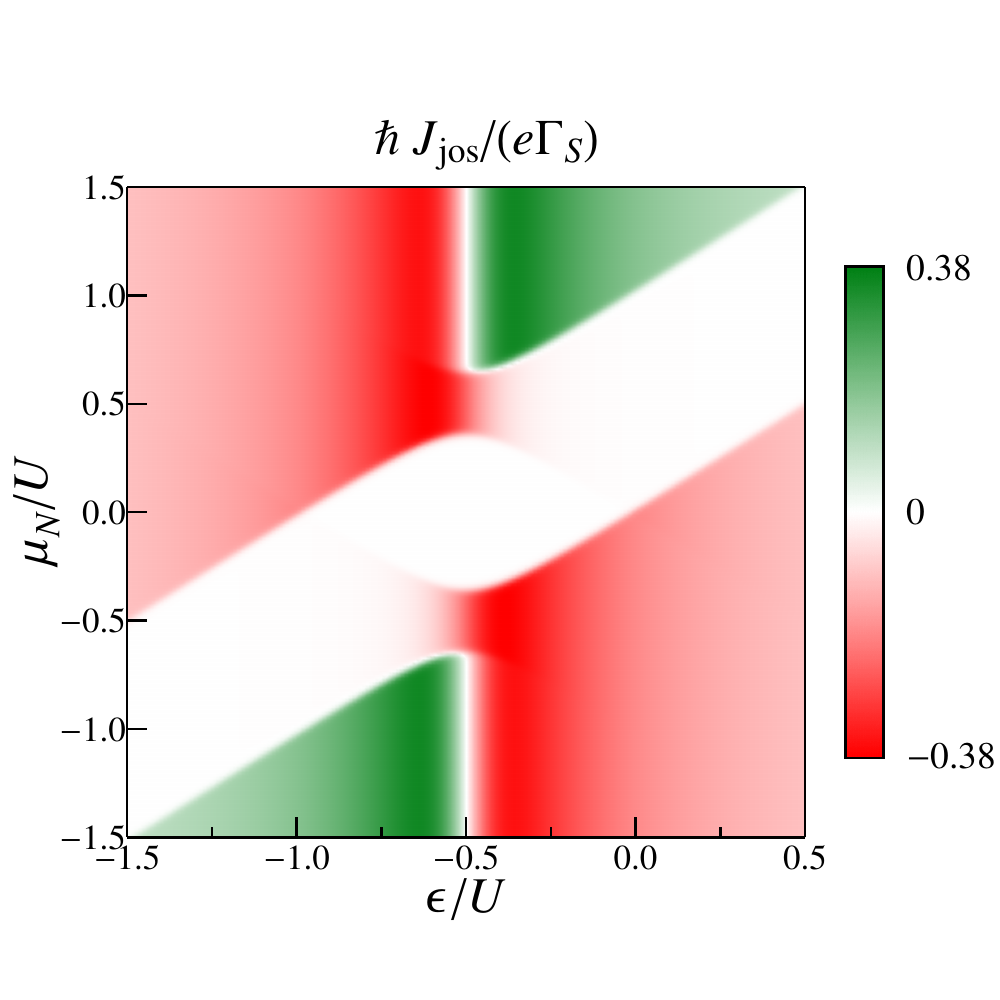}
	\caption{(Color online) Density plot of the Josephson current in the limit of $\Delta\rightarrow\infty$ as a function
	 of the level position $\epsilon$ and the chemical potential of the normal lead $\mu_{N}$. The other parameters are $\Gamma_{S}=0.2 U$, 
	 $\Gamma_{N}=0.001 U$, $\Phi=\pi/2$, and $k_{B}T=0.01 U$.  
\label{density-plot-Josephson}}
\end{figure}

\begin{figure}
	\includegraphics[width=0.9\columnwidth]{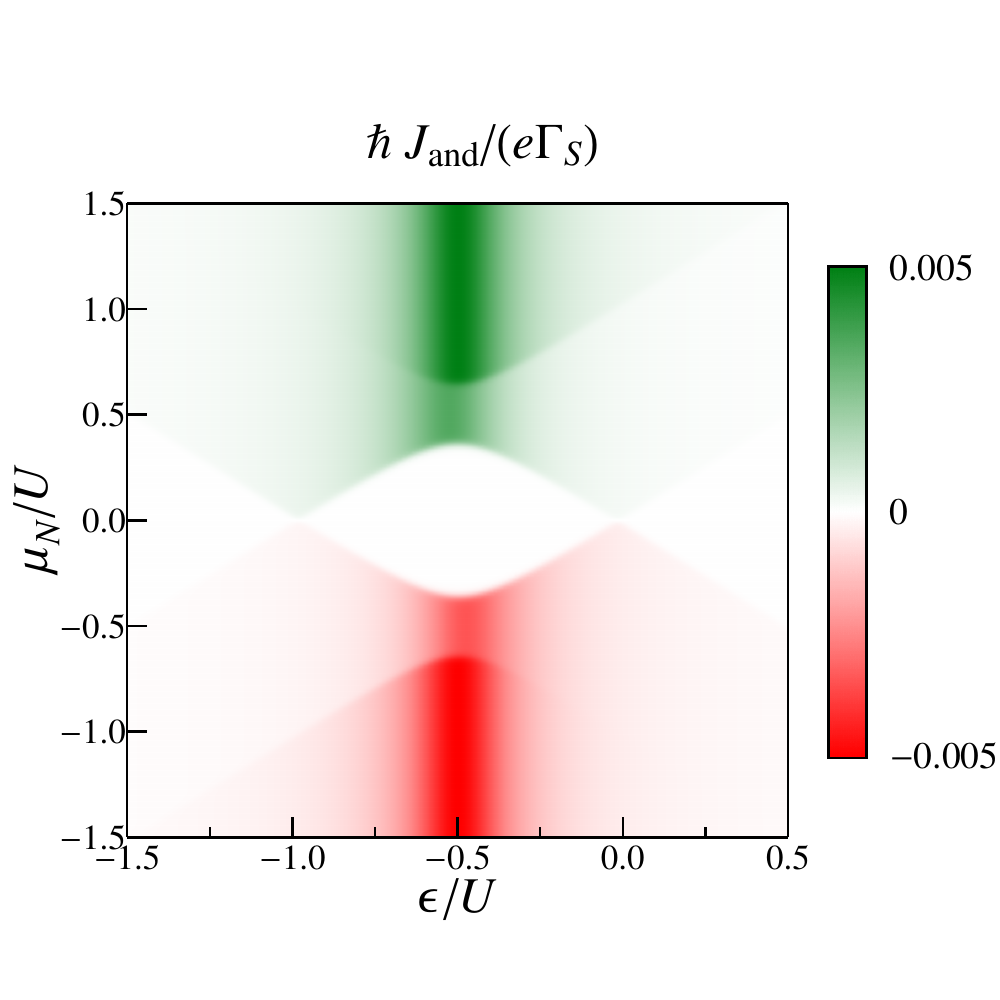}
	\caption{(Color online) Density plot of the Andreev current in the limit of $\Delta\rightarrow\infty$ as a function of the
	level position $\epsilon$ and the chemical potential of the normal lead $\mu_{N}$. The other parameters are $\Gamma_{S}=0.2 U$, $\Gamma_{N}
	=0.001 U$, $\Phi=\pi/2$, and $k_{B}T=0.01 U$.  
\label{density-plot-Andreev}}
\end{figure}

We assume symmetric tunnel couplings to the superconducting leads, $\Gamma_L=\Gamma_R=\Gamma_S$.
Furthermore, we choose a gauge such that $\Phi_L=-\Phi_R=\Phi/2$.
For both the Josephson and the Andreev current we calculate the value for the $\Delta\rightarrow\infty$ limit and the correction to first order in $1/\Delta$. 
The tunnel coupling to the normal lead, $\Gamma_N$ is accounted for in lowest non-vanishing order, i.e., to zeroth order for the Josephson current and to first order for the Andreev current.

\begin{figure}
	\includegraphics[width=0.9\columnwidth]{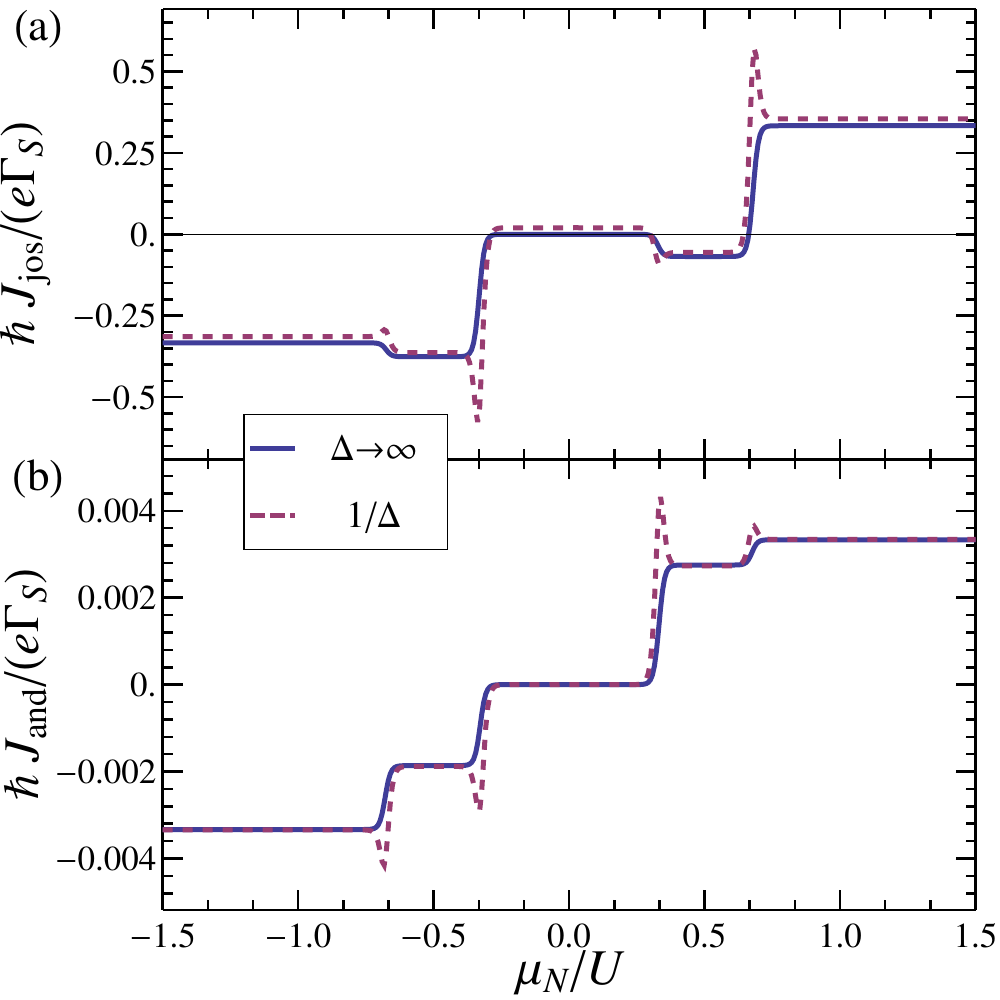}
	\caption{(Color online) Plot of (a) the Josephson current and (b) the Andreev current as a function the chemical potential of the normal lead $\mu_{N}$.
	The solid (blue) curve shows the $\Delta\rightarrow\infty$ limit, the dashed (red) curve includes the first $1/\Delta$ correction.
	The other parameters are $\epsilon=-0.4U$, $\Gamma_{S}=0.2 U$, $\Gamma_{N}=0.001 U$, $\Delta=5U\;$, $\Phi=\pi/2$, and $k_{B}T=0.01 U$.  
\label{JCombined}}
\end{figure}

Figures \ref{density-plot-Josephson} and \ref{density-plot-Andreev} show the Josephson current and the Andreev current in the limit of $\Delta \rightarrow \infty$ as a function of the level position $\epsilon$ and the chemical potential of the normal lead $\mu_{N}$.
This is the limit considered in Ref.~\onlinecite{governale08} and is used as a reference for the present discussion.
A cut through the density plots at a fixed level position $\epsilon=-0.4U$ is shown in Fig.~\ref{JCombined} (solid blue line).
As the $1/\Delta$ expansion is expected to be valid for $\Delta$ being the dominating energy scale with respect to other system parameters such as the charging energy $U$, the tunnel-coupling strengths $\Gamma_{\eta}$ or the level position $\epsilon$, we have chosen in Fig.~\ref{JCombined} a rather large value of $\Delta=5U$.
The currents display a step-like behavior with the position of the steps reflecting the ABS energies. 
This is compared with the currents expanded up to first order in $1/\Delta$, see dashed red lines in Fig.~\ref{JCombined}.
From this comparison, we can draw two conclusions.

First, away from the ABS, the solid and dashed curves almost coincide, i.e., the $1/\Delta$ correction to the current is small.
This leads to the conclusion that away from the ABSs the $\Delta\rightarrow\infty$ calculations provide a useful and quantitatively accurate description of sub-gap transport also for finite values of the superconducting gap.

Second, we find peaks at the positions of the ABSs in the $1/\Delta$ correction to the current.
While the peaks themselves are artefacts of the $1/\Delta$ expansion, they carry important physical information about the renormalization of the ABSs as compared to their $\Delta\rightarrow\infty$ value.
This can be understood in the following way.
Suppose we knew the exact expression for the current at finite $\Delta$. It is natural to assume that steps in the current as function of bias voltage occur at the ABSs for finite $\Delta$ and that these steps stem from Fermi functions $f(E_{A,\gamma,\gamma'})$ that appear in the diagrammatic language after integrating over all energies $\omega_i$.
A systematic $1/\Delta$ expansion of the exact expression for the current would include the expansion of $E_{A,\gamma,\gamma'}$ inside the Fermi function, giving rise to derivatives of the Fermi function at the $\Delta\rightarrow \infty$ value of the ABSs, $f'(E_{A,\gamma,\gamma'}^{\Delta\rightarrow\infty})$.
These appear as peaks in the current at the position of the infinite-gap ABSs, as it is the case in Fig.~\ref{JCombined}.
The sign of the peaks relative to the sign of the steps indicates the direction of the renormalization.
The ABS is renormalized towards (away from) zero if step and peak have the same (the opposite) sign.
In the examples plotted in Fig.~\ref{JCombined} (a) and (b) the curves indicate a renormalization of all four ABSs towards zero.

\subsection{Renormalization of Andreev Bound States}
\label{level-renormalization}

The $1/\Delta$ expansion indicates that the ABSs are renormalized as compared to the $\Delta\rightarrow\infty$ limit.  
It even predicts the sign of the renormalization.
It does, however, not provide a possibility to calculate the ABSs at finite $\Delta$.
For this task, alternative approaches are required.
In the limit of vanishing Coulomb interaction, the problem can be solved exactly.\cite{beenakker95,splettstoesser07}
A mean-field treatment of the Coulomb interaction provides easily an estimate of the ABSs.\cite{pillet10,martin-rodero12}
Numerical renormalization group (NRG) calculations are exact but numerically demanding.\cite{yoshioka00,bauer07,hecht08,lim08,martin-rodero12}
Instead, we propose a resummation approach that is more accurate than mean field but  less 
computationally heavy than NRG.

In the following, we consider the limit of weak tunnel coupling to the normal lead, i.e., we discuss the spectrum to zeroth order in $\Gamma_N$.
Furthermore, we choose $\Phi=0$ without loss of generality (to consider the case of a finite $\Phi$ one has to replace $\Gamma_S$ by $\Gamma_S \cos (\Phi/2)$; to model the case of a single superconducting lead, $\Gamma_S$ needs to be replaced by $\Gamma_S/2$).

\subsubsection{Exact Green's function for the noninteracting dot}
\label{noninteracting-dot}
\begin{figure}
	\includegraphics[width=0.9\columnwidth]{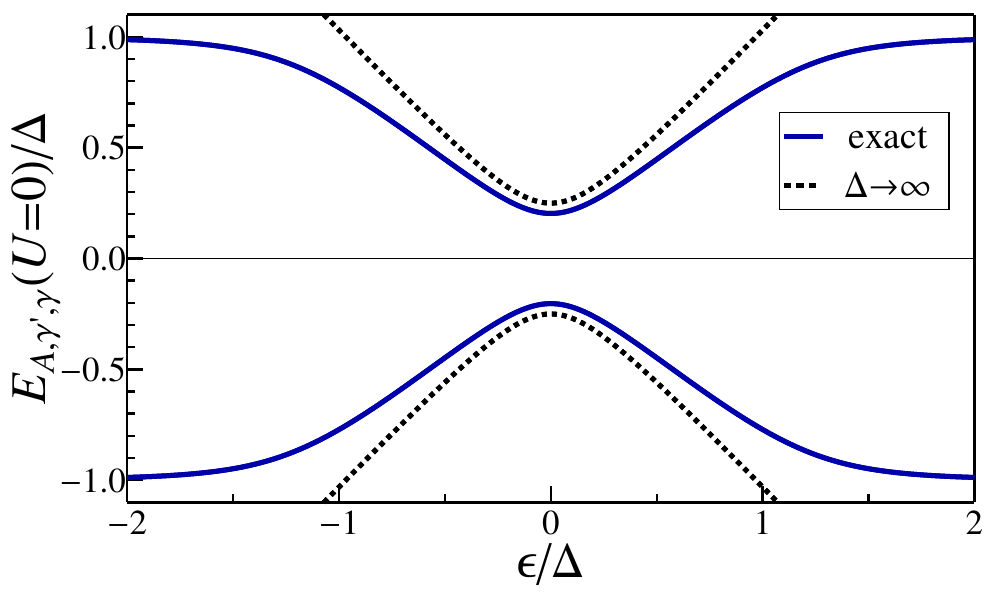}
	\caption{(Color online) Plot of the $U=0$ Andreev-bound states for a tunnel-coupling strength to the superconducting lead of 
	$\Gamma_S=0.25\Delta$ as a function of the level position $\epsilon$. 
	All energies are normalized to $\Delta$.
	The solid (blue) line shows the exact result, the dashed (black) line the bound states computed by means of the $\Delta\rightarrow\infty$ expression, which in the non-interacting case is simply $\pm \epsilon_A$. 
\label{ABS-zeroU-plot}}
\end{figure}

For a noninteracting dot, $U=0$, the exact Green's function for spin $\sigma$ is given, in Nambu space,  by 
\begin{equation}
\label{exact-greens-function}
\hat{G}^{r}=\left( \omega \cdot \mathbf{I}_{2\times 2} -\hat{h}_{\sigma}(U=0)-\Sigma_{S} \right)^{-1},
\end{equation}
with $\mathbf{I}_{n\times n}$ being the identity matrix in $n$ dimensions.
The matrix
\begin{equation}
\hat{h}_{\sigma}(U=0)=
	\begin{pmatrix}
		\epsilon_{\sigma} & 0\\
		0 & -\epsilon_{-\sigma}
	\end{pmatrix}
\end{equation}
accounts for the single-particle energies, and the self-energy due to tunneling reads
\begin{equation}
	\hat{\Sigma}_{S}= \Gamma_S
	\begin{pmatrix}
		-\frac{\omega}{\sqrt{\Delta^2-\omega^2}} &  \frac{\Delta}{\sqrt{\Delta^2-\omega^2}} \\
		  \frac{\Delta}{\sqrt{\Delta^2-\omega^2}} & - \frac{\omega}{\sqrt{\Delta^2-\omega^2}} 
	\end{pmatrix}.
\end{equation}
The ABSs, i.e. the excitation energies, are found as the real part of the poles of the Green's function's determinant.
An example is shown in Fig.~\ref{ABS-zeroU-plot}.
In the noninteracting case, there are only two ABSs.
Their energies for finite $\Delta$ are reduced as compared to the $\Delta\rightarrow\infty$ limit.
In particular, the curvature of the energy as a function of the bare level position $\epsilon$ changes for large $|\epsilon|$, such that the ABSs lie within the window $[-\Delta,\Delta]$.

\subsubsection{Hartree-Fock approximation}
\label{HFA}
A simple and quick method to include Coulomb interaction is Hartree-Fock (HF). 
In self-consistent HF treatments,\cite{shiba73,rozhkov99,yoshioka00,karrasch08,martin-rodero12} the dot occupations have to be calculated self-consistently. 
An even simpler treatment of the Coulomb interaction has been used in Ref.~\onlinecite{pillet10} to fit the experimental data.
This approach corresponds to replace $\hat{h}_{\sigma}(U=0)$ in Eq.~(\ref{exact-greens-function}) by
\begin{equation}
	\hat{h}_{\sigma}(U)=
	\begin{pmatrix}
		\tilde{\epsilon}_{\sigma} & 0\\
		0 & -\tilde{\epsilon}_{-\sigma}
	\end{pmatrix},
\end{equation}
where $\tilde{\epsilon}_{\downarrow}=\tilde{\epsilon}_{\uparrow}+U$, i.e. spin symmetry is broken by hand.
For $U=0$, the exact Green's function Eq.~\eqref{exact-greens-function} is recovered.
Furthermore, it reproduces the exact ABSs in the $\Delta\rightarrow\infty$ limit.
In the following, we use this simpler version of HF for comparison with the results from the resummation approach.

\subsubsection{Resummation Approach}
\label{resum}

An important feature of the diagrammatic technique presented in this paper is that Coulomb interaction can be included beyond the mean-field level.
The downside is that for an exact solution one has to sum up infinitely many diagrams.
A great simplification is achieved by working in the basis $\{\uparrow, \downarrow, +,- \}$.
As discussed above, in the $\Delta\rightarrow\infty$ limit, all diagrams that contain superconducting lines are cancelled by cross vertices.
This means that all contributing diagrams contain normal tunnel lines only. 

At finite $\Delta$, the situation is different.
Now, both superconducting lines and cross vertices dress the original normal tunnel lines.
We define the propagator $\Pi(\omega)$ as the diagram part between two vertices that are connected by a normal tunnel line with energy $\omega$ running from right to left.
This propagator obeys a Dyson equation
\begin{equation}
\label{dyson-equation}
	\Pi(\omega) = \Pi^{(0)}(\omega)+\Pi(\omega) \; \Sigma(\omega) \; \Pi^{(0)}(\omega),
\end{equation}
where $\Pi^{(0)}(\omega)$ is the free propagator  (without superconducting lines and cross vertices) and $\Sigma(\omega)$  the self energy.
 
\begin{figure}
	\includegraphics[width=0.45\columnwidth]{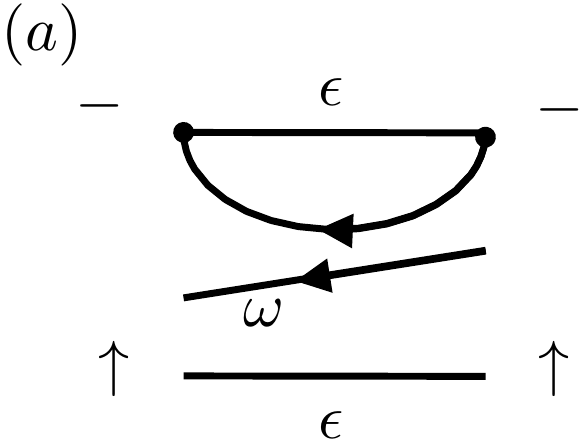} \qquad
	\includegraphics[width=0.45\columnwidth]{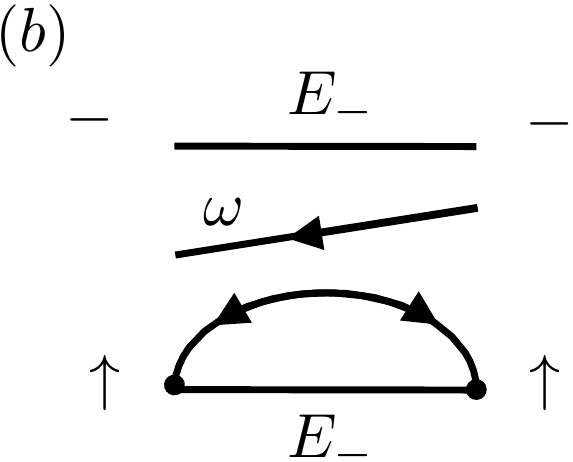}
	\caption{Examples for diagrams that contribute to the self-energy.
\label{resum-diagrams}}
\end{figure}

In the $\Delta\rightarrow\infty$ limit, the self energy is exactly zero. 
For an approximative treatment of the case of finite $\Delta$, we now include all self energy parts that contain one superconducting tunnel line or one cross vertex.
Examples are shown in Fig.~\ref{resum-diagrams}.

Since the Hilbert space is four dimensional, the dimension of the corresponding Liouville space and, thus, of the matrices appearing in Eq.~(\ref{dyson-equation}) is, in general, 16. 
However, as a consequence of spin and charge conservation while tunneling, the Dyson equation decouples into blocks of $4\times 4$ matrices.
The presence of the normal tunneling line running from right to left which carries a definite charge and spin puts constraints on the possible states at the beginning and end of the upper ($\chi_1$) and lower ($\chi_2$) contour.
In the following, we choose, without loss of generality, $(\chi_1, \chi_2) \in \{ (-,\uparrow) , (\downarrow,+) , (\downarrow,-) , (+,\uparrow) \}$, which is the relevant case when the normal tunneling line carries spin $\downarrow$.
Then, 
\begin{eqnarray}
	&&\left(\Pi^{(0)}(\omega)\right)^{-1} = \nonumber \\
	&&\omega \cdot \mathbf{I}_{4\times 4} - \mathrm{diag}(E_{A,+,-},E_{A,-,-},E_{A,-,+},E_{A,+,+}) 
\end{eqnarray}
for the inverse free propagator, and the self energy
\begin{equation}
\Sigma(\omega) = 
	\begin{pmatrix}
		W^{--}_{\uparrow \uparrow}(\omega) & W^{-\downarrow}_{\uparrow +}(\omega) & W^{-\downarrow}_{\uparrow -}(\omega) & W^{-+}_{\uparrow \uparrow}(\omega)\\
		\\
		W^{\downarrow -}_{+\uparrow}(\omega) & W^{\downarrow \downarrow}_{++}(\omega) & W^{\downarrow\downarrow}_{+-}(\omega) & 
		W^{\downarrow +}_{+\uparrow}	(\omega)\\
		\\
		W^{\downarrow -}_{-\uparrow}(\omega) & W^{\downarrow \downarrow}_{-+}(\omega) & W^{\downarrow\downarrow}_{--}(\omega) & 
		W^{\downarrow +}_{-\uparrow}	(\omega)\\
		\\
		W^{+-}_{\uparrow\uparrow}(\omega) & W^{+\downarrow}_{\uparrow +}(\omega) & W^{+\downarrow}_{\uparrow -} (\omega) & W^{++}_{\uparrow\uparrow}(\omega)
	\end{pmatrix},
\end{equation}
is given by the generalized rates $W^{\chi_1 \chi_{1'}}_{\chi_2 \chi_{2'}}(\omega)$ that contain a left-going external line with energy $\omega$.

We solve Eq.~\eqref{dyson-equation} for the full propagator to obtain
\begin{equation}
	\Pi(\omega)=\left[ \left( \Pi^{(0)}(\omega) \right)^{-1} -\Sigma(\omega) \right]^{-1}.
\end{equation}
The excitation energies of the proximized dot are probed by the external energy $\omega$ and they are given by the real poles of the full propagator.
As mentioned above, we approximate the self-energy $\Sigma(\omega)$ by including only diagrams that contain a single superconducting tunneling line or one cross vertex, in addition to the external line of energy $\omega$, see Fig.~\ref{resum-diagrams}.

Although the resummation scheme does not define a controlled approximation, its results are with a few limitations quantitatively accurate, as we discuss in the following.
It is clear that, by construction, the $\Delta\rightarrow\infty$ limit is reproduced exactly. 
Also for the noninteracting case, $U=0$, the resummation approach yields the exact solution.
As an artifact, however, the resummation scheme produces for $U=0$ two extra, unphysical, solutions in addition to the two correct ones shown in Fig.~\ref{ABS-zeroU-plot}.

\begin{figure}
	\includegraphics[width=\columnwidth]{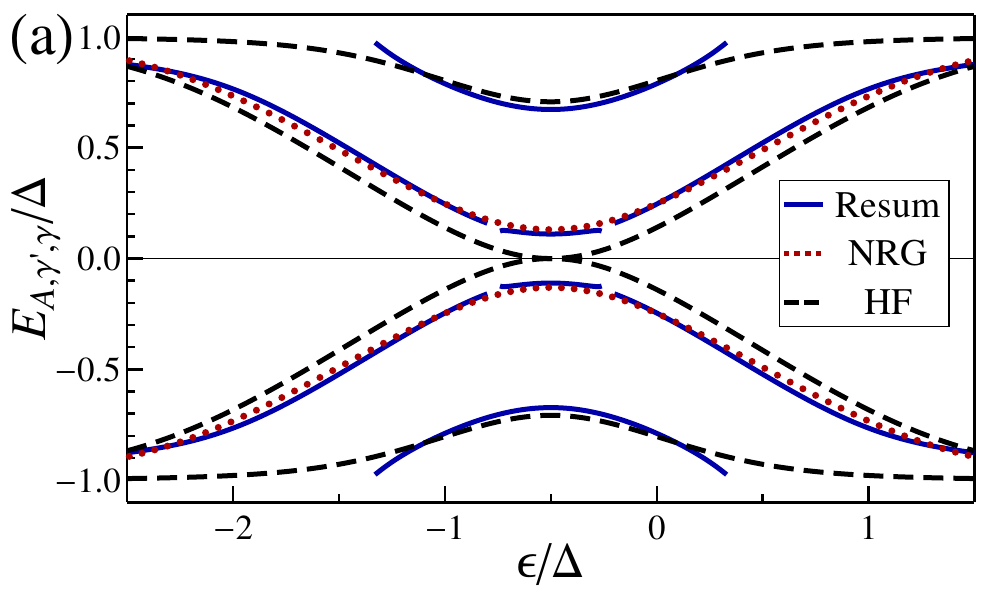}
	\includegraphics[width=\columnwidth]{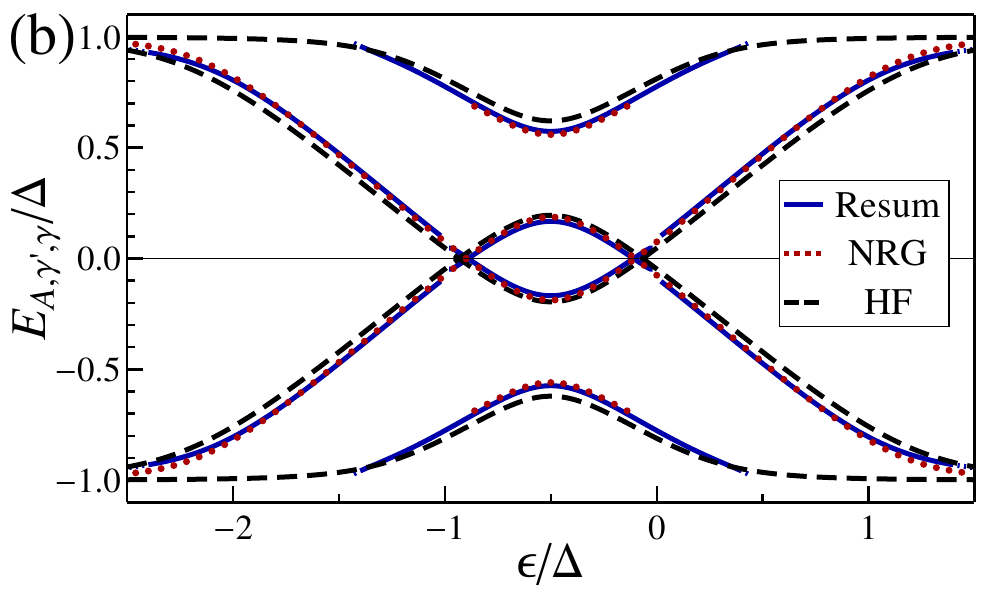}
	\includegraphics[width=\columnwidth]{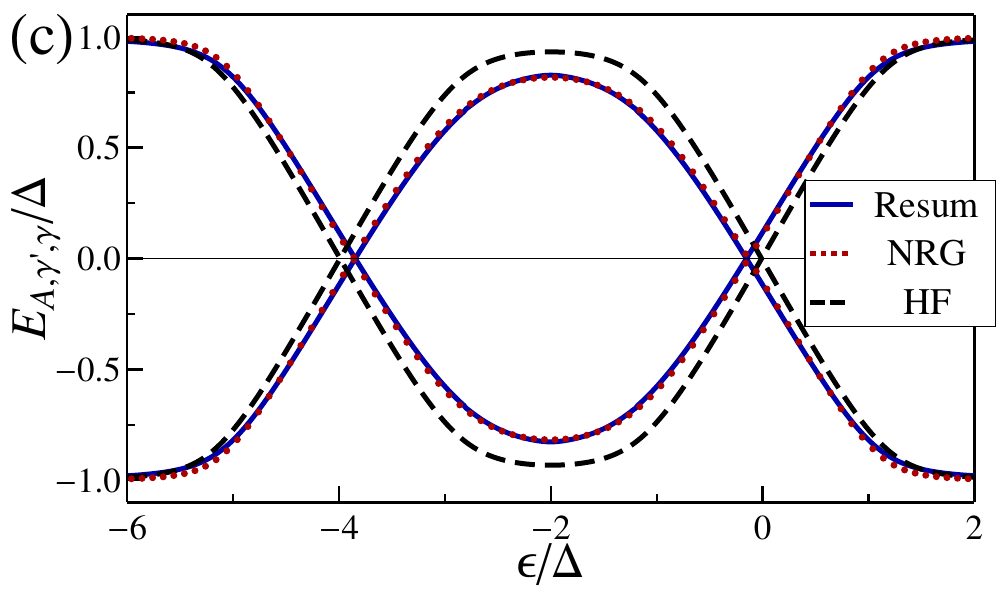}
	\caption{(Color online) Plot of the Andreev-bound states from the resummation approach, NRG\cite{NRG-footnote}, and the Hartree-Fock approximation as a function of the
	level position $\epsilon$ for (a) $U=\Delta$ and $\Gamma_S=0.5\Delta$, (b) $U=\Delta$ and $\Gamma_S=0.2\Delta$, and (c) $U=4\Delta$ and $\Gamma_S=0.2\Delta$. 
	\label{ABS-plot}}
\end{figure}

The most interesting regime, however, is the case of finite $U$ and finite $\Delta$.
Results obtained from the resummation approach for different values of $\Gamma_S$ and $U$ are presented in Fig.~\ref{ABS-plot} as solid (blue) curves.
Depending on the values of the parameters of the model, we find either two or four solutions.
The ABSs are renormalized as compared to the $\Delta\rightarrow\infty$ solution.
The sign of the renomalization agrees with the prediction of the $1/\Delta$ expansion of Sec.~\ref{finite-gap-expansion}.
For comparison, we also show the results from HF as dashed (black) lines. 
While qualitatively similar, there are substantial quantitative deviations.
First, HF seems to underestimate the renormalization of the ABSs.
Second, the position of the crossing points of the ABS energies in the resummation scheme is shifted as compared to the HF and the $\Delta\rightarrow\infty$ result, see Fig.~\ref{ABS-plot}(c).

A reliable quality check of the proposed approximation, however, is only given by the comparison with the full NRG results.\cite{NRG-footnote}
We find a remarkably good agreement between NRG data, dotted (red) lines in Fig.~\ref{ABS-plot}, and resummation approach.
The ABS spectrum is quantitatively reproduced in almost the entire parameter space.
The main difference is that the resummation scheme sometimes yields four solutions when NRG only predicts two.
Furthermore, sometimes the resummation approach produces solutions with finite imaginary part.
Since the ABSs in the absence of coupling to a normal reservoir  are sharp  due to the gap in the quasiparticle spectrum, these solutions are discarded. This gives rise to the small gaps in the solid (blue) line in 
Fig.~\ref{ABS-plot}(a) and (b).
Both features (extra pair of solutions and poles with finite imaginary part)
are artifacts of the approximation employed.

Finally, we remark that in the present context we use the resummation approach only to calculate the excitation spectrum of dot plus superconducting leads.
This is an equilibrium quantity that can also be addressed within genuine equilibrium techniques.
In fact, a self-consistent renormalization approach to describe ABSs at finite $\Delta$ within a Matsubara formalism has been put forward in Ref.~\onlinecite{meng09}.
The resummation approach based on the real-time diagrammatic technique, on the other hand, is not restricted to equilibrium situations but can also be used to describe the current in the nonequilibrium regime.

\section{Beyond weak coupling to the normal lead}
\label{results-2nd-ord-Gn}

In the previous section we have discussed how a finite superconducting gap in the leads renormalizes the ABSs of a quantum dot weakly tunnel-coupled to a normal lead.
Now, we want to address the influence of the tunnel coupling to the normal lead beyond the weak-coupling limit.
For this, we go back to $\Delta\rightarrow \infty$ but include next-to-leading order corrections in $\Gamma_N$.
To distinguish the contributions of different order in $\Gamma_N$, we introduce in the following an index $(n)$ for $n$-th order.
We study the Josephson and the Andreev current as well as the average charge of the quantum dot. 
Within a perturbation expansion, they start to zeroth, first, and zeroth order, respectively, i.e., the next-to-leading order corrections are of first, second, and first order, respectively.
To evaluate them we have to include diagrams with two tunneling lines from the normal lead.
This increases the number of diagrams considerably and we generate and evaluate them by means of a computer code.
At this point, we would like to emphasize that although the Josephson current starts in zeroth order in the coupling to the normal lead, it still depends on the chemical potential of the normal lead for the following reason.
According to Eq.~\eqref{current-formula}, the zeroth-order Josephson current is given as a product of the zeroth-order current rates and the zeroth-order elements of the reduced density matrix.
To calculate the latter, it is necessary to expand the generalized master equation, see Eq.~\eqref{master-equation}, up to first order in $\Gamma_N$, so that the chemical potential of the normal lead enters the zeroth-order density-matrix elements and, thus, the zeroth-order Josephson current.

It is natural for a systematic perturbation expansion in a small parameter that higher-order corrections are quantitatively small.
Therefore, we focus, in the following, on qualitative features that appear in the region close to the point of zero detuning $\delta=2\epsilon+U=0$ when going beyond the weak-coupling limit.
Around zero detuning, the weights of the states $|0\rangle$ and $|d\rangle$ entering the coherent superpositions $|\pm\rangle$ become equal, indicating a large proximization of the quantum dot.
For bias voltages large enough such that the quantum dot has a finite probability to be either empty or doubly occupied, 
Josephson and Andreev currents set in.
To lowest order in the tunnel coupling to the normal lead, zero detuning defines a resonance condition between dot and superconductors, visible in the Josephson current, the Andreev current, and the average charge.
The inclusion of next-order corrections shifts the resonance condition which can be interpreted as a renormalization of the energy difference between empty and doubly occupied dot.

\begin{figure}
	\includegraphics[width=\columnwidth]{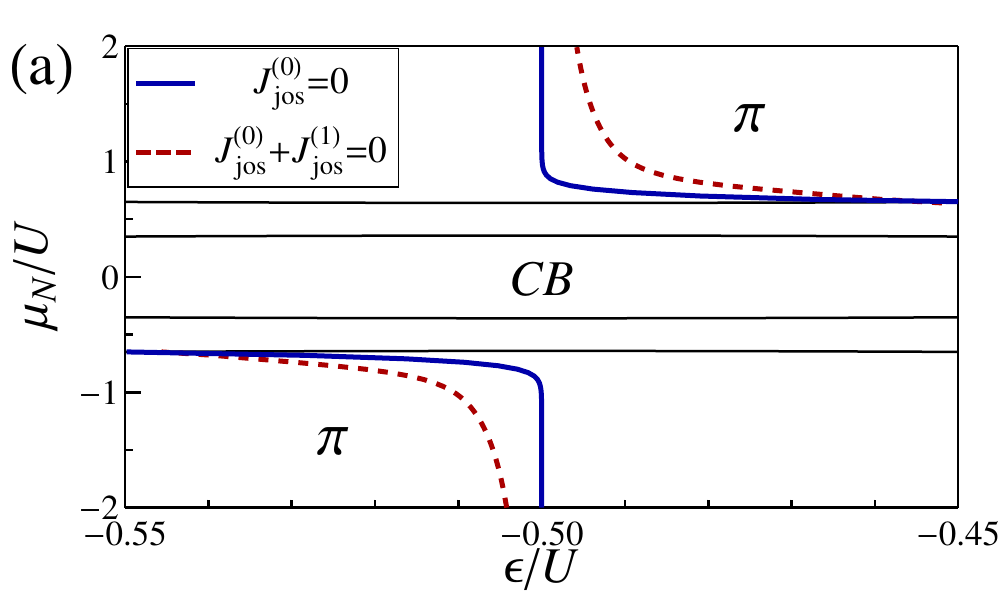}
	\includegraphics[width=\columnwidth]{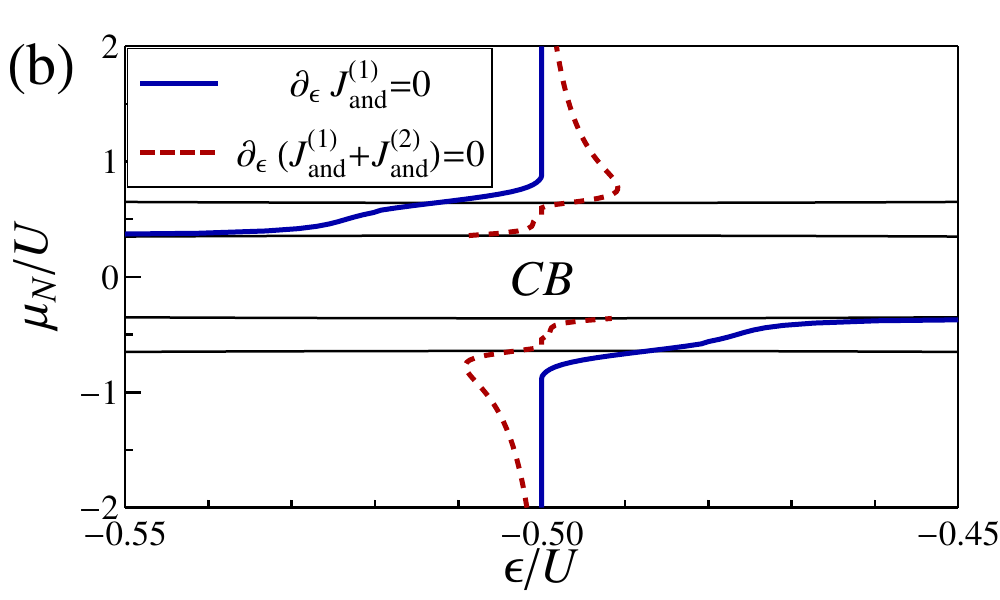}
	\caption{(Color online) Plot of (a) the position of the $0-\pi$~transition of the Josephson current and (b) the extreme values of the Andreev current.
	The chosen parameters are $k_B T/U=0.05$, $\Gamma_N/U=0.05$, $\Gamma_S/U=0.2$, and $\Phi=\pi/2$.
	In (a) the $\pi$~junctions are indicated by the symbol $\pi$.
	The thin black lines in both plots show the position of the ABSs and $CB$ marks the Coulomb-blockaded area between the two inner ABSs.
\label{renormalizations}}
\end{figure}

In the regime of large bias voltage, the Josephson current changes sign as function of the quantum dot's level position $\epsilon$.
This indicates a transition from a $0$- to a $\pi$-junction behavior. 
In the limit $\Delta\rightarrow\infty$ and weak tunnel coupling to the normal lead,\cite{governale08} this $0-\pi$~transition happens exactly at zero detuning, $\epsilon=-U/2$, and then bends when the bias voltage approaches the highest ABS, see Fig.~\ref{density-plot-Josephson}.
When including the next-order correction in $\Gamma_N$, the position of this $0-\pi$~transition becomes renormalized, see Fig.~\ref{renormalizations}~(a).
The origin of this renormalization can be understood in terms of an effective, tunnel-coupling induced field that acts on the isospin introduced in Eq.~(\ref{isospin_definition}).
A finite bias voltage leads to the generation of the $z$-component of the isospin. 
The tunnel coupling to the superconductors leads to an effective field along the $x$-direction.
This causes the isospin to rotate and acquire a finite $y$-component.
According to Eq.~(\ref{josephson_iso}), however, a finite $x$-component of the isospin is needed for a Josephson current.
This is accomplished with an additional finite $z$-component of the effective field.
To lowest order, this component is given by the detuning, $B_z^{(0)}=\delta$, which explains the position of the $0-\pi$~transition in Fig.~\ref{density-plot-Josephson}.
The next-order correction, $B_z^{(1)}$, renormalizes this position.

We remark that this renormalization of the $0-\pi$~transition has already been predicted in Ref.~\onlinecite{pala07}, in which the tunnel couplings to both the superconductors and the normal leads as well as the detuning $\delta$ have been simultaneously treated as small parameters, i.e., the systematic perturbation expansion included already both $B_z^{(0)}$ and $B_z^{(1)}$.
On the other hand, the formation of ABSs could not be treated in that expansion.
In Ref.~\onlinecite{governale08}, on the other hand, no constraints were put on $\Gamma_S$ and $\delta$, i.e., the ABSs could be described, but the renormalization $B_z^{(1)}$ did not contribute to lowest order in $\Gamma_N$.
Only the present calculation, with arbitrary $\Gamma_S$ and $\delta$ and next-to-leading order on $\Gamma_N$ enables us to address both the ABSs and the renormalization of the position of the $0-\pi$~transition.

While the Josephson current changes sign at small detuning, the Andreev current becomes extremal as a function of $\epsilon$, see Fig.~\ref{density-plot-Andreev}.
In Fig.~\ref{renormalizations}~(b) we show the position of this extremum in the $\epsilon$-$\mu_N$-plane.
(We do not show the position of the extremum in the Coulomb-blockade regime since there the current is exponentially suppressed). 
To lowest order and for large bias voltage, the extremum is at zero detuning. 
The next-order correction, however, leads to a renormalization of this position, for the same reason as the renormalization of the $0-\pi$~transition in the Josephson current.
 
Finally, we address the average quantum dot charge $Q$, which is related to the $z$-component of the isospin via
$Q = -e\left( 1+2 I_z \right)$.
Again, we focus on the region of small detuning where the proximity effect is most pronounced.
In the Coulomb-blockade regime (at small bias voltage), the dot is preferably singly occupied, $Q=-e$.
A larger bias voltage has the tendency to inject a second electron into the quantum dot.
Superconducting correlations, however, mix the states of empty and double occupation and, thus, reduce the average number of electrons.
For zero detuning $\delta$ and to zeroth order in $\Gamma_N$, the charge $Q$ remains $-e$ even for large bias voltage.
Only when increasing $|\delta|$ the number of charges increases above $1$.
In the intermediate-bias regime for finite detuning values the average charge tends towards $0$ or $-2e$, depending on the sign of the detuning.
This effect can easily be understood by looking at the eigenstates $|\pm\rangle$:
in the intermediate-bias regime for positive values of $\mu_N$, i.e. $E_{A,+,-}<\mu_N<E_{A,+,+}$, the state $|-\rangle$ can already be excited while state $|+\rangle$ is still inaccessible.
From the definition of state $|-\rangle$ it follows that a positive (negative) detuning causes the weight of the zero component to dominate over (be dominated by) the double component resulting in an decrease (increase) of the average number of electrons.

\begin{figure}
	\includegraphics[width=\columnwidth]{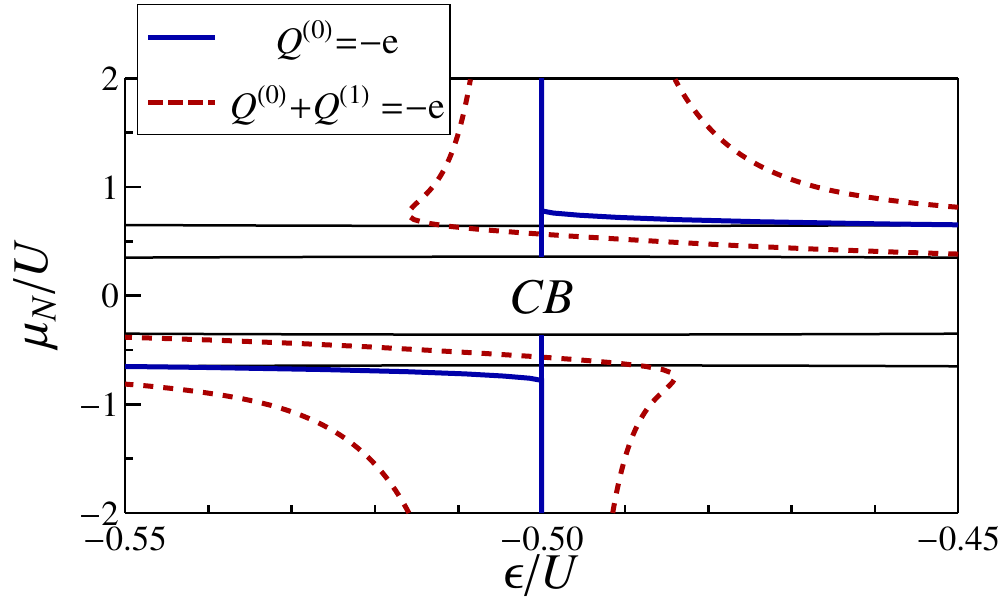}
	\caption{(Color online) Plot of the average dot charge. The chosen parameters are $k_B T/U=0.05$, $\Gamma_N/U=0.05$, $\Gamma_S/U=0.2$, and $\Phi=\pi/2$.
	The thin black lines show the position of the ABSs and $CB$ marks the Coulomb-blockaded area between the two inner ABSs.
	\label{AverageDotCharge}}
\end{figure}

In Fig.~\ref{AverageDotCharge} we show the position of $Q/(-e)=1$, or equivalently $I_z=0$, for both zeroth and first order.
(We do not plot the position in the Coulomb-blockade regime.)
As discussed above, the position of $Q^{(0)}/(-e)=1$ is at zero detuning in the large-bias regime and splits in the intermediate-bias regime, with $Q^{(0)}/(-e)<1$ between and $Q^{(0)}/(-e)>1$ outside the two lines.
When including the next-order correction, the situation changes. 
Now, the line of $[Q^{(0)}+Q^{(1)}]/(-e)=1$ is split even for the large-bias regime, opening a region with $[Q^{(0)}+Q^{(1)}]/(-e)<1$ in between.
To understand this, we analyze $I_z$ at zero detuning, $\delta=0$.
Solving the generalized master equation\cite{governale08,footnote} immediately yields $I_z^{(0)}\left( \delta=0 \right) = 0$ and $I_z^{(1)}\left( \delta=0 \right) = -A_y^{(1)}/B_x^{(0)}$, where $B_x^{(0)} = 2 |\chi|$ and
\begin{eqnarray*}
A_y^{(1)} &=& \frac{1}{4} \; \mathrm{Im}\left[ W_{+-}^{-- {(1)}}+W_{++}^{-+ {(1)}}+W_{--}^{+- {(1)}}+W_{-+}^{++ {(1)}} \right], \\
&=& \frac{\Gamma_N}{4 \pi} \sqrt{1-\frac{\delta^2}{4\epsilon_{A}^{2}}} \sum_{\gamma,\gamma'=\pm} \mathrm{Re}\left[ \psi\left( \frac{1}{2} + i \frac{E_{A,\gamma',\gamma} -\mu_N}{2\pi k_B T} \right) \right] .
\end{eqnarray*}
Here $W_{\chi_2 \chi'_2}^{\chi_1 \chi'_1{(1)}}$ are generalized rates in first order in $\Gamma_N$ and $\psi(x)$ is the digamma function.
It is, thus, the isospin generation term $A_y^{(1)}$, describing combined Andreev processes that involve both the normal and the superconducting leads, which reduce $Q/(-e)$ below $1$.

\section{Conclusions}
\label{sec:concl}

The theoretical description of sub-gap (Josephson and Andreev) transport through quantum dots with strong Coulomb interaction coupled to normal and superconducting leads simplifies substantially in the limit of infinitely-large superconducting gap $\Delta$ and weak tunnel coupling $\Gamma_N$ to the normal lead.
For experimental devices with finite $\Delta$ and larger $\Gamma_N$ outside the Kondo regime, these calculations may still be used as an approximation.
The assessment of the quality of this approximation to describe sub-gap transport has been the focus of this paper. 
In particular we came to the following conclusions.
The position of the ABSs strongly depend on $\Delta$, i.e., a $\Delta\rightarrow\infty$ approximation is insufficient.
These ABSs define threshold voltages in the current-voltage characteristics at which new transport channels open.
The numerical values for the Josephson and Andreev current between these threshold voltages, however, seem to be nicely approximated by the $\Delta\rightarrow\infty$ limit.

Similarly, we find that the next-to-leading order correction in the tunnel coupling $\Gamma_N$ to the normal lead mainly yields small quantitative corrections.
But there are also qualitative differences: the position of the $0-\pi$~transition in the Josephson current and the peak positions in the Andreev current are shifted.

In conclusion, we find that the $\Delta\rightarrow\infty$ and weak-coupling calculations provide a very good approximation for sub-gap transport with the limitation that the positions of the ABSs for finite $\Delta$ are not properly described.
To address the latter, we proposed a resummation approach that substantially improves over mean-field treatments and favorably compares with more elaborate NRG calculations.
 
\acknowledgements
D.F. appreciates the hospitality of The Victoria University of Wellington. Financial support from DFG via KO 1987/5 is gratefully acknowledged.
We are grateful to A. Mart\'{i}n-Rodero and A. Levy Yeyati for fruitful discussions and for sending us their NRG data.
We would like to thank A. Hucht and V. Meden for helpful discussions.


\begin{thebibliography}{99}


%Review articles
\bibitem{franceschi10}
S. DeFranceschi, L. Kouwenhoven, C. Sch\"onenberger, and W. Wernsdorfer, Nat. Nanotech. {\bf 1}, 703 (2010).

\bibitem{martin-rodero11}
A. Mart\'{i}n-Rodero and A. Levy Yeyati, Adv. Phys. \textbf{60}, 899 (2011).

% Andreev reflection
\bibitem{degennes63} 
P.-G. de Gennes and D. Saint-James, Phys. Lett. \textbf{4}, 151 (1963).
\bibitem{andreev64} 
A. F. Andreev, Zh. Eksp. Teor. Fiz. {\bf 46}, 1823 (1964); [Sov. Phys.
JETP {\bf 19}, 1228 (1964)].


%Experiments
\bibitem{buitelaar02}
M. R. Buitelaar, T. Nussbaumer, and C. Sch\"onenberger, Phys. Rev. Lett. \textbf{89}, 256801 (2002).
\bibitem{buitelaar03}
M. R. Buitelaar, W. Belzig, T. Nussbaumer, B. Babi\'c, C. Bruder, and C. Sch\"onenberger, Phys. Rev. Lett. \textbf{91}, 057005 (2003).
\bibitem{cleuziou06}
J.-P. Cleuziou, W. Wernsdorfer, V. Bouchiat, T. Ondar\c{c}uhu, and M. Monthioux, Nature Nanotechnology \textbf{1}, 53 (2006).
\bibitem{jarillo-herrero06}
P. Jarillo-Herrero, J. A. van Dam, and L. P. Kouwenhoven, Nature \textbf{439}, 953 (2006).
\bibitem{ingerslev06}
H. I J{\o}rgensen, K. Grove-Rasmussen, T. Novotn\'y, K. Flensberg, and P. E. Lindelof, Phys. Rev. Lett. \textbf{96}, 207003 (2006).
\bibitem{grove-rasmussen07}
K. Grove-Rasmussen, H. Ingerslev J{\o}rgensen, and P. E. Lindelof, New J. Phys. \textbf{9}, 124 (2007).
\bibitem{ingerslev07}
H. Ingerslev J{\o}rgensen, T. Novotn\'y, K. Grove-Rasmussen, K. Flensberg, and P. E. Lindelof, Nano Lett. \textbf{7}, 2441 (2007).
\bibitem{eichler07}
A. Eichler, M. Weiss, S. Oberholzer, C. Sch\"onenberger, A. Levy Yeyati, J. C. Cuevas, and A. Mart\'in-Rodero, Phys. Rev. Lett. \textbf{99}, 126602 (2007).
\bibitem{sand-jespersen07}
T. Sand-Jespersen, J. Paaske, B. M. Andersen, K. Grove-Rasmussen, H. I. J{\o}rgensen, M. Aagesen, C. B. S{\o}rensen, P. E. Lindelof, K. Flensberg, and J. Nyg{\aa}rd, Phys. Rev. Lett. \textbf{99}, 126603 (2007).
\bibitem{eichler09}
A. Eichler, R. Deblock, M. Weiss, C. Karrasch, V. Meden, C. Sch\"onenberger, and H. Bouchiat, Phys. Rev. B \textbf{79}, 161407(R) (2009).
\bibitem{deacon10}
R. S. Deacon, Y. Tanaka, A. Oiwa, R. Sakano, K. Yoshida, K. Shibata, K. Hirakawa, and S. Tarucha, Phys. Rev. Lett. \textbf{104}, 076805 (2010).
\bibitem{deacon10prb}
R. S. Deacon, Y. Tanaka, A. Oiwa, R. Sakano, K. Yoshida, K. Shibata, K. Hirakawa, and S. Tarucha, Phys. Rev. B \textbf{81}, 121308(R) (2010).


%Exp. - pi-transition and beam splitters
\bibitem{vandam06}
 J. A. van Dam, Y. V. Nazarov, E. P. A. M. Bakkers, S. De Franceschi, and L. P. Kouwenhoven, Nature \textbf{442},667 (2006).
 \bibitem{hofstetter09}
 L. Hofstetter, S. Csonka, J. Nyg{\aa}rd, and C. Sch\"onenberger, Nature \textbf{461}, 960 (2009).
\bibitem{herrmann10}
L. G. Herrmann, F. Portier, P. Roche, A. Levi Yeyati, T. Kontos, and C. Strunk, Phys. Rev. Lett. \textbf{104}, 026801 (2010).
\bibitem{das12}
A. Das, Y. Ronen, M. Heiblum, D. Mahalu, A.V. Kretinin, H. Shtrikman, Nat. Commun. \textbf{3}, 1165 (2012).

 %Exp. ABS
\bibitem{pillet10}
J.-D. Pillet, C. Quay, P. Morfin, C. Bena, A. L. Yeyati, and P. Joyez, Nature Phys. \textbf{6}, 965 (2010).
\bibitem{dirks11}
T. Dirks, T. L. Hughes, S. Lal, B. Uchoa, Y.-F. Chen, C. Chialvo, P. M. Goldbart, and N. Mason, Nature Phys. \textbf{7}, 386 (2011).

% Greensfunction for U=0
\bibitem{beenakker95}
C. W. J. Beenakker, in: \textit{Mesoscopic Quantum Physics}, edited by E. Akkermans, G. Montambaux, J.-L. Pichard, and J. Zinn-Justin (North-Holland, Amsterdam, 1995): pp. 279-324.
 
%theo. finite gap, U approximated
\bibitem{fazio98}
 R. Fazio and R. Raimondi, Phys. Rev. Lett. \textbf{80}, 2913 (1998).
\bibitem{schwab99}
P. Schwab and R. Raimondi, Phys. Rev. B \textbf{59}, 1637 (1999).
\bibitem{clerk00}
A. A. Clerk, V. Ambegaokar, and S. Hershfield, Phys. Rev. B \textbf{61}, 3555 (2000).
\bibitem{cuevas01}
J. C. Cuevas, A. Levy Yeyati, and A. Mart\'in-Rodero, Phys. Rev. B \textbf{63}, 094515 (2001).
\bibitem{avishai03}
Y. Avishai, A. Golub, and A. D. Zaikin, Phys. Rev. B \textbf{67}, 041301 (2003).
\bibitem{dell-anna08}
L. Dell`Anna, A. Zazunov, and R. Egger, Phys. Rev. B \textbf{77}, 104525 (2008).
\bibitem{koerting10}
V. Koerting, B. M. Andersen, K. Flensberg, and J. Paaske, Phys. Rev. B \textbf{82}, 245108 (2010).


%theo. perturbation in Gamma
\bibitem{pala07}
M. G. Pala, M. Governale, and J. K\"onig, New. J. Phys. \textbf{9}, 278 (2007).
\bibitem{governale08}
M. Governale, M. G. Pala, and J. K\"onig, Phys. Rev. B \textbf{77}, 134513 (2008).


%infinite gap
\bibitem{rozhkov00}
A. V. Rozhkov and D. P. Arovas, Phys. Rev. B \textbf{62}, 6687 (2000).
\bibitem{tanaka07}
Y. Tanaka, A. Oguri, and A. C. Hewson, New J. Phys. \textbf{9}, 115 (2007).
\bibitem{karrasch08}
C. Karrasch, A. Oguri, and V. Meden, Phys. Rev. B \textbf{77}, 024517 (2008).
\bibitem{futterer09}
 D. Futterer, M. Governale, M. G. Pala, and J. K\"onig, Phys. Rev. B \textbf{79}, 054505 (2009).
\bibitem{sothmann10}
B. Sothmann, D. Futterer, M. Governale, and J. K\"onig, Phys. Rev. B \textbf{82}, 094514 (2010).
 \bibitem{futterer10}
 D. Futterer, M. Governale, and J. K\"onig, Europhys. Lett. \textbf{91}, 47004 (2010).
\bibitem{eldridge10}
J. Eldridge, M. G. Pala, M. Governale, and J. K\"onig, Phys. Rev. B \textbf{82}, 184507 (2010).
\bibitem{hiltscher11}
B. Hiltscher, M. Governale, J. Splettstoesser, and J. K\"onig, Phys. Rev. B \textbf{84}, 155403 (2011).
\bibitem{braggio11}
A. Braggio, M. Governale, M.G. Pala, and J. K\"onig, Solid State Commun. \textbf{151}, 155 (2011).
\bibitem{moghaddam12}
A.G. Moghaddam, M. Governale, and J. K\"onig, Phys. Rev. B \textbf{85}, 094518 (2012).
\bibitem{hiltscher12}
B. Hiltscher, M. Governale, and J. K\"onig, Phys. Rev. B \textbf{86}, 235427 (2012).

%multiple Andreev reflection
\bibitem{yeyati97}
A. Levy Yeyati, J. C. Cuevas, A. L\'{o}pez-D\'{a}valos, and A. Mart\'{i}n-Rodero, Phys. Rev. B \textbf{55}, R6137 (1997).
\bibitem{johansson99}
G. Johansson, E. N. Bratus, V. S. Shumeiko, and G. Wendin, Phys. Rev. B \textbf{60}, 1382 (1999).
\bibitem{sun02}
Q. F. Sun, H. Guo, and J. Wang, Phys. Rev. B \textbf{65}, 075315 (2002).
\bibitem{yeyati03}
A. Levy Yeyati, A. Mart\'{i}n-Rodero, and E. Vecino Phys. Rev. Lett. \textbf{91}, 266802 (2003).
\bibitem{jonckheere09}
T. Jonckheere, A. Zazunov, K. V. Bayandin, V. Shumeiko, and T. Martin, Phys. Rev. B \textbf{80}, 184510 (2009).

%NRG vs HF
\bibitem{martin-rodero12}
A. Mart\'{i}n-Rodero and A. Levy Yeyati, J. Phys.: Condens. Matter \textbf{24}, 385303 (2012).

%Kondo
\bibitem{graeber04}
M. R. Gr\"aber, T. Nussbaumer, W. Belzig, and C. Sch\"onenberger, Nanotechnology \textbf{15}, S479 (2004).
\bibitem{hofstetter10}
L. Hofstetter, A. Geresdi, M. Aagesen, J. Nyg{\aa}rd, C. Sch\"onenberger, and S. Csonka, Phys. Rev. Lett. \textbf{104}, 246804 (2010).
\bibitem{cho99}
S. Y. Cho, K. Kang, and C.-M. Ryu, Phys. Rev. B \textbf{60}, 16874 (1999).
\bibitem{sun01}
Q.-F. Sun, H. Guo, and T.-H. Lin, Phys. Rev. Lett. \textbf{87}, 176601 (2001).
\bibitem{aono03}
T. Aono, A. Golub, and Y. Avishai, Phys. Rev. B \textbf{68}, 045312 (2003).
\bibitem{siano04}
F. Siano and R. Egger, Phys. Rev. Lett. \textbf{93}, 047002 (2004).
\bibitem{krawiec04}
M. Krawiec and K. I. Wysoki\'{n}ski, Supercond. Sci. Technol. \textbf{17}, 103 (2004).
\bibitem{splettstoesser07}
J. Splettstoesser, M. Governale, J. K\"onig, F. Taddei, and R. Fazio, Phys. Rev. B \textbf{75}, 235302 (2007).
\bibitem{tanaka07-2}
Y. Tanaka, N. Kawakami, and A. Oguri, J. Phys. Soc. Jpn. \textbf{76}, 074701 (2007).
\bibitem{domanski08}
T. Doma\'{n}ski and A. Donabidowicz, Phys. Rev. B \textbf{78}, 073105 (2008).
\bibitem{domanski08-2}
T. Doma\'{n}ski, A. Donabidowicz, and K. I. Wysoki\'{n}ski, Phys. Rev. B \textbf{78}, 144515 (2008).
\bibitem{yamada10}
Y. Yamada, Y. Tanaka, and N. Kawakami, J. Phys. Soc. Jpn. \textbf{79}, 043705 (2010).
\bibitem{yamada11}
Y. Yamada, Y. Tanaka, and N. Kawakami, Phys. Rev. B \textbf{84}, 075484 (2011).

\bibitem{footnote-ABS}
A slightly more precise denotation would be {\it Andreev excitation energies}. For simplicity, however, we use the term {\it Andreev bound state energy} throughout the paper.

\bibitem{yoshioka00}
T. Yoshioka and Y. Ohashi, J. Phys. Soc. Japan {\bf 69}, 1812 (2000).

\bibitem{bauer07}
J. Bauer, A. Oguri, and A.C. Hewson, J. Phys.: Condens. Matter {\bf 19}, 486211 (2007).

\bibitem{hecht08}
T. Hecht, A. Weichselbaum, J. von Delft, and R. Bulla, J. Phys.: Condens. Matter {\bf 20}, 275213 (2008).

\bibitem{lim08}
J.S. Lim and M.S. Choi, J. Phys.: Condens. Matter {\bf 20}, 415225 (2008).

%self-consistent HF
\bibitem{shiba73}
H. Shiba, Prog. Theor. Phys. \textbf{50}, 50 (1973).
\bibitem{rozhkov99}
A. V. Rozhkov and D. P. Arovas, Phys. Rev. Lett. \textbf{82}, 2788 (1999).



%footnote
\bibitem{NRG-footnote}
The NRG data, some of them presented in Ref.~\onlinecite{martin-rodero12}, were kindly provided by A. Mart\'{i}n-Rodero and A. Levy Yeyati.

%finite-gap ABS
\bibitem{meng09}
T. Meng, S. Florens, and P. Simon, Phys. Rev. B \textbf{79}, 224521 (2009).

%footnote
\bibitem{footnote}
Please note that in Ref. \onlinecite{governale08} $A_y^{(1)}$ has mistakenly been taken to be zero. For Ref. \onlinecite{governale08} this error has no influence on the results because all investigated quantities have been independent of $A_y^{(1)}$.

\end{thebibliography}
\end{document}